\documentclass{article}
\usepackage{epsfig}
\usepackage{amssymb}
\usepackage{amsmath}
\usepackage{amsfonts}
\newcommand \be{\begin{eqnarray}}
\newcommand \ee{\end{eqnarray}}
\begin{document}
\begin{center}
{\bf Nuclear Saturation and Correlations.}\\
\bigskip
\bigskip

H. S. K\"ohler \footnote{e-mail: kohler@physics.arizona.edu} \\
{\em Physics Department, University of Arizona, Tucson, Arizona
85721,USA} \\
S. A. Moszkowski \footnote{e-mail: stevenmos@ucla.edu}\\
{\em UCLA, Los Angeles, CA 90095,USA}\\
\end{center}
\date{\today}

\begin{abstract}
The relation between nuclear saturation and NN-correlations is
examined. Nucleons bound in a nucleus have a reduced effective mass
due to the mean field. This results in off-energy-shell scatterings
modifying the free-space NN-interaction by a dispersion correction. 
This is a major contribution to the density-dependence of the effective
in-medium force and
to saturation. Low-momentum effective interactions have been derived by 
renormalisation methods whereby correlations may be reduced by
effectively cutting off high momentum components of the interaction. 
The effect of these cut-offs on dispersive corrections
and on saturation is the main focus of this paper.
The role of the
tensor-force, its strength and its effect on correlations is of particular
interest. The importance of the definition of the mean field
in determining saturation as well as  compressibility is also pointed
out.

With a cut-off below $\sim 2.6 fm^{-1}$ there is no saturation
but at lower density the binding energy is still well approximated 
suggesting that such a force may be useful in nuclear structure
calculations of (small) finite nuclei if saturation is not an issue.

A separable interaction that fits experimental
phase-shifts exactly by inverse scattering methods is used. Recent
experiments measure short ranged correlations (SRC's) to be $0.23$ for
$^{56}Fe$. 
Other experiments have obtained a depletion of occupation-numbers in
$^{208}Pb$ to be $\sim 0.2$.
For nuclear matter with the separable interaction and a
continuous spectrum we obtain the related quantity 
$\kappa$ to be $0.175$ with the Bonn-B deuteron parameters, while
Machleidt's  gets  $\kappa= 0.125$ for the Bonn-B potential and a
continuous spectrum.

\end{abstract}

\section{Introduction}
The problem of saturation of nuclear forces is nearly as old as nuclear
physics but is not yet satisfactorily resolved theoretically. 
The first efforts to explain the saturation in terms of nuclear forces
was the combination of a nonexchange (Wigner) and a space-exchange
(Majorana) part. It is found that  a Majorana exchange of four times the
strength of the Wigner part is required.\cite{bet36,bro67} This disagrees with
experimental scattering data. 
These early studies assumed a "well-behaved" potential such that
perturbation theory could be used. The subsequent finding that the
$^{1}S_{0}$ phase-shift changes sign at about 250 $MeV$ led
Jastrow\cite{jas51} to
propose a very strong short ranged repulsion, a "hard core". 
This it seemed would also help to explain the saturation property of
nuclear matter at least qualitatively. A short ranged
repulsion although soft rather than hard has since been part of all 
modern potentials and is theoretically justified.

The strong nature of the nuclear forces does however exclude a low order 
perturbation
approach as the short-ranged repulsion leads to strong short ranged
correlations. The way out of this dilemma was led by Brueckner some 50
years ago developping the theory bearing his name. 
His approach to solving nuclear many-body problems has dominated 
all aspects  of nuclear physics since, be it nuclear structure or
nuclear matter. The problem of nuclear saturation is \it qualitatively
\rm solved applying this theory with modern nuclear forces. The
saturation is found to result from a combination of the properties of
the two-body nucleon force (being repulsive at short distances, being
state-dependent and having a tensor-component) and many body effects
(correlation effects related to  Pauli and dispersion effects). 
It is generally accepted that the contribution from three-body forces
is also important.

Although the problem of the saturation of nuclear forces 
may be partially resolved  with experimentally and theoretically well
motivated two- and three- body forces
it is still unclear whether the discrepancies found are due to incomplete
knowledge of the "free" nuclear forces, the effective interactions in
nuclei, nucleonic degrees of freedom or relativistic effects etc.
The major topic of the present work is related to our incomplete
knowledge of the short-range details of the NN-force as opposed to the
long-ranged pion-exchange part. In the "new" approach to the nuclear
many-body problem the short-ranged part is 'integrated out' with the goal of 
reducing the many body problem to that of a low- (or even first-) order
problem and eliminating the need of Brueckner or related many-body
techniques.
In the present work we show the effect on correlations and on saturation 
as a result of the typical cut-offs in
momentum space suggested by this approach. 

We use Brueckner's theory of nuclear matter for our calculations. The
central part of this theory is the
Reaction matrix $K$ (we use Brueckner's original notation $K$ rather
than the often used $G$ to avoid confusion with Green's function),
defined by
\begin{equation}
K=V+V{Q\over{e}}K
\label{K}
\end{equation}
with $V$ being the "free" NN-interaction,
$Q/e$ the in-medium propagator and 
where $Q$ is the Pauli-operator and $e$ is a sum of kinetic and
self-consistent mean field energies
\begin{equation}
U(k)=\sum_{k'<k_{F}}<kk'|K|kk'>
\label{U}
\end{equation}
The calculations below use this definition of $U$ for both hole and
particle states usually referred to as  the "continuous" choice.
The Brueckner expression for the total energy is given by
\begin{equation}
E=\sum_{k<k_{F}} { k^{2}\over {2m}}+{1\over{2}}
\sum_{k',k<k_{F}}<kk'|K|kk'>
\label{E}
\end{equation}
The many-body effects of Brueckner's theory 
are clearly expressed in the Moszkowski-Scott
separation method \cite{mos60,hsk61}.  One particularly important
effect as regards
saturation is the dispersion correction.
$K$ will have off-energy shell contributions due
to the momentum dependence of the mean field binding $U(k)$ and this is the
origin of this correction.

Let $\Delta U$ be the
average change in potential energy in  intermediate states when solving
eq. (\ref{K}).
The dispersion correction to the $K$-matrix
is obtained by differentiating $K$ in eq.  (\ref{K})
by the energy denominator $e$ to get \cite{mos60}
\begin{equation}
\Delta K_{disp} \propto \Delta U*I_{w}.
\label{disp}
\end{equation}
where the wound-integral $I_{w}$ is defined  by
$$I_{w}=\int (\Psi(r)- \Phi(r))^{2}d{\bf r}$$
with $\Psi$ and $\Phi$  the {\bf correlated} and {\bf uncorrelated}
two-body  wave-functions respectively.
The dispersion correction to the energy per particle $E/A$ is then
\begin{equation}
\Delta E_{disp}/A  \propto \Delta U*\kappa
\label{dispE}
\end{equation}
where $\kappa=\rho*I_{w}$ with $\rho$ being the density.

The dispersion term is small at low density (small
finite nuclei) but grows with density because of the increased binding.
It is repulsive and is therefore an important
contribution to saturation. It is basically a three-body effect
as the effective two-body interaction depends on the mean field due to
the presence of "third nucleons" that constitute the mean field $U$. 
The wound-integral $I_{w}$ is an important quantity by itself being a
measure of the correlation strengths.

A good understanding of the origin of the dispersion correction  
is of particular interest with the present "new" approach to the
nuclear many-body problem using EFT or similar ideas. In the $V_{low-k}$
approximation  with $k < 2-3 fm^{-1}$,
short-ranged correlations are \it \'a priori \rm ignored and
a nuclear matter calculation gives no saturation, unless supplemented
with a three-body interaction of not well-defined origin. 
In the EFT-approach a low-energy effective interaction is derived but
a three-body force is simultaneously generated and may contribute to
saturation. This 
three-body force is however of different origin than the above mentioned
dispersion-effect which is a medium or many-body-effect. 
It is of course now generally accepted that the \it bona fide \rm
three-body force is also large enough to be an important factor in 
understanding the saturation.

The main subject of this paper is to look at
the effect
of typical momentum cut-offs related to EFT and $V_{low-k}$ and 
how it relates to two-body correlations in
nuclear matter and saturation calculations.
In addition we address the more general problem of relation between 
saturation and nuclear forces.

Section II shows the main results of the numerical work. Part A
deals with the dispersion corrections while Part B is concerned with the
related depletion factor $\kappa$ with some results summarized in a
Table. It is well-known that tensor correlations in nuclei are important
but that they are also to some extent not well determined. We deal
with this subject in Section III. In Section IV we discuss some topics
related to higher order corrections such as due to the spectral widths
and hole-hole propagation. 
A short Summary and Conclusions are found in Section V.

\section{Numerical results}
The separable potential derived from inverse
scattering was calculated 
as described in  previous papers
\cite{kwo95,hsk04}. The Brueckner calculation was done as in numerous
previous papers by one of us.  The effective interaction $K(\omega,k,P)$
in eq. (\ref{K}) was calculated 
as a function of the three variables, the starting energy $\omega$,
relative momentum $k$ and center of mass momentum $P$. 
The Pauli-operator in eq. (\ref{K}) was a function of the two variables
$k$ and $P$, the  angle-averaged approximation.
The sum of the mean fields $U$ in the energy-denominator
$$e=\omega-2k_{i}^{2}-U({\bf P}/2+{\bf k_{i}})-U({\bf P}/2-{\bf k_{i}})$$
of eq. (\ref{K})
was approximated  by
$$2*U((P^{2}/4+k_{i}^{2})^{1\over{2}})$$ which implies a qudratic
approximation of the mean field around each value of $P/2$. 
Here $k_{i}$ are summed over when solving eq. (\ref{K}).
The $P$
contribution to the kinetic energy is cancelled in $e$.
$U(k)$ was calculated from eq.(\ref{U}) by summing ${\bf k}'$ over the
fermisea. This involved integrating over the angle between ${\bf k}$ and
${\bf k'}$ while $P$ and the starting energy $\omega$ are functions of this
angle. The mean field contribution to the starting energy
$\omega=2k^{2}+
U(k)+U(k')$ was calculated as above in an effective mass approximation
to get $\omega=2k^{2}+2*U({1\over{2}}(k^{2}+k'^{2}))$. The total energy in eq.
(\ref{E}) was calculated by summing over the mean field $U(k)$.

With present day computing power some of the approximations used above
are not necessary. Improved calculations show considerable corrections
\cite{fri02}.

\subsection{Dispersion effect}

\begin{figure}
\centerline{
\psfig{figure=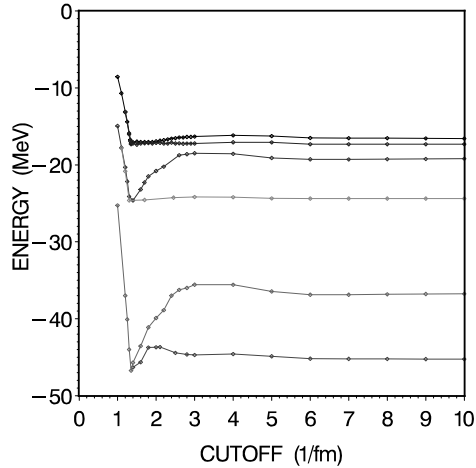,width=7cm,angle=0}
}
\vspace{.0in}
\caption{
Effects of the selfconsistent mean field (dispersion-correction).
There are three sets of curves. The uppermost set shows the contribution to 
the potential energy per particle  from the
$^{1}S_{0}$ state, the middle from the $^{3}S_{1}$ and the bottom
includes all (21) states. In each set of curves the lower curve is without
the mean field $U(k)$ while the upper is with $U(k)$ included
in the calculation. The difference between these two curves is
the dispersion correction, which is seen  to decrease as the
cutoff $\Lambda$ decreases below $\Lambda \sim 3.0 fm^{-1}$ and
approaches zero as $\Lambda \rightarrow k_{F}=1.35 fm^{-1}$.
}
\label{satfig1}
\end{figure}

The importance of the dispersion effect was emphasized in the
Introduction.
In this regard we show Fig. \ref{satfig1} which has
three sets of graphs, each consisting of two curves. 

\begin{figure}
\centerline{
\psfig{figure=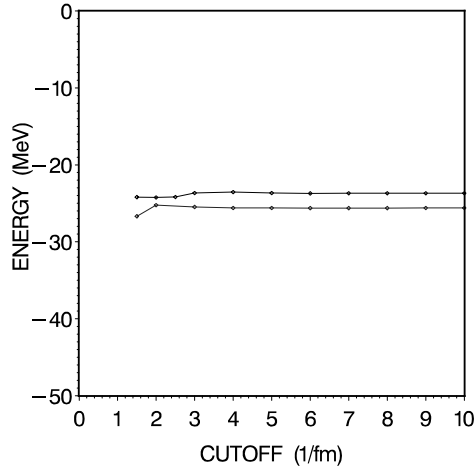,width=7cm,angle=0}
}
\vspace{.0in}
\caption{
The $^{3}S_{1}$ contribution to the potential energy per particle
without the tensor force. The upper curve is with and the lower is without
the mean field. It is seen that the difference, the dispersion
correction, is small without the tensor force.
Compare with the middle set of curves in Fig. \ref{satfig1}.
The density is here given by $k_{F}=1.35 fm^{-1}$
}
\label{satfig2}
\end{figure}

The upper
curve of each set shows the binding energy per particle as a function of
cutoff $\Lambda$, including
dispersion, while the lower is without dispersion. The no dispersion
curves are the result of Brueckner calculation without the selfconsistent
mean field $U$ ( i.e. with $e \rightarrow e_{0}$) in eq (\ref{K}) but
with the $Q$-operator.
The uppermost set of curves shows the $^{1}S_{0}$ contribution to the
energy per particle while the middle is for the $^{3}S_{1}$ and the
lowest set includes all (21) states. We point out that the no-dispersion
results are practically independent of cutoff, while the repulsive
dispersion results decrease below about $3 fm^{-1}$ to join the
no-dispersion result at $\Lambda=k_{F}$. From the discussion
above this serves to show that the effect of the 
\it correlations \rm between the nucleons is
decreasing for cutoffs below $3 fm^{-1}$. The origin of this effect will
be discussed further in Section B.
This result should also be compared with the $\approx 2 fm^{-1}$ 
where the phase-shifts turn repulsive.
Note also that the dispersion
correction for the $^{1}S_{0}$ state is appreciably smaller than for the
$^{3}S_{1}$ state. This difference is to be attributed to the 
tensor correlations.
This is further illustrated by Fig. \ref{satfig2} which, when compared  
with Fig. \ref{satfig1} shows 
that the short-ranged correlations in the $^{3}S_{1}$-state
contribute relatively little to the total dispersion correction. The
dominant correlations are due to the tensor force. Fig. \ref{satfig1}
also shows that the dispersions and correlations due to the other states
(beyond the $S$-states)
are not negligible.

\begin{figure}
\centerline{
\psfig{figure=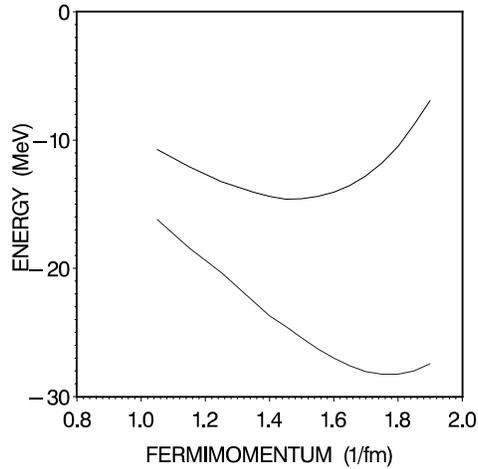,width=7cm,angle=0}
}
\vspace{.0in}
\caption{
The upper curve is a  Brueckner calculation of the total
energy per particle as a function of the fermi-momentum. The lower curve
shows the result without the selfconsistent mean field U but only
kinetic energies (i.e. with $e
\rightarrow e_{0}$) in eq.( \ref{K}) and therefore no
dispersion-correction, while the only many-body effect comes from the
Pauli-operator $Q$.
}
\label{satfig3}
\end{figure}

 Fig. \ref{satfig3}  shows the importance of the
dispersion correction in providing saturation in a Brueckner calculation
of the binding energy. The separable interaction without any cut-off
is used here. The upper curve is
the full Brueckner calculation, while in the lower the selfenergy $U(k)$
is neglected so that the only many-body effect comes from the
$Q$-operator..

\begin{figure}
\centerline{
\psfig{figure=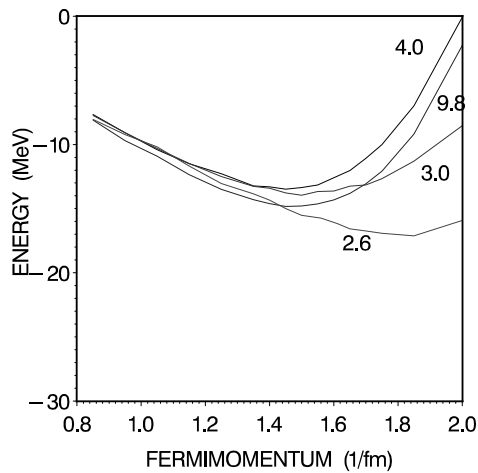,width=7cm,angle=0}
}
\vspace{.0in}
\caption{
These curves show the effect of high momentum cut-off on the saturation
curve. These are Brueckner calculations of the total energy per
particle as a function of fermi-momentum for the indicated  values of the
cut-off momentum $\Lambda$ in units of $fm^{-1}$.
}
\label{satfig4}
\end{figure}

The effect of the high momentum cut-off is further illustrated by Fig.
\ref{satfig4}. With $\Lambda=9.8 fm^{-1}$  the phase-shifts for all
available energies are
included in calculating the separable potential
while with decreasing $\Lambda$ these are also decreased in energy
accordingly. (See ref \cite{hsk04}).
For all the indicated values of $\Lambda$ the
binding energies around and below the experimental saturation density are
approximately equal. At higher densities it is however 
seen that for the smallest
value shown here, i.e. $\Lambda=2.6 fm^{-1}$ there is essentially no  saturation.
(For $\Lambda=2 fm^{-1}$ there is of course even less evidence of
saturation as shown in Fig. \ref{br1st2nd}.)
The effect of 
the correlations decreases with $\Lambda$ and the dispersion correction 
at the higher densities becomes eventually 
too small to give saturation at a reasonable density. This lack of
saturation was also shown in ref \cite{hsk04} and agrees also
qualitatively with Fig. 1 of ref \cite{boz06}.  It will be further 
discussed below in Sect. B in relation to  the depletion factor $\kappa$.

\subsection{Depletion factor $\kappa$}
The (short-ranged) correlations cause scatterings to states outside 
the fermisea. This results in the depletion of the normally occupied
(model) states.\cite{dah69,hsk69,hsk93} This is
quantified by  $\kappa$ that we define as (see e.g. refs.
\cite{haf70,bet71})
\begin{equation}
\kappa_{i}={1\over 8}(2J+1)(2T+1)\rho*I_{w}
\label{kappa}
\end{equation}
where the wound-integral $I_{w}$ was defined above and $\rho$ is the
density.
$\kappa$ is state-dependent as is $I_{w}$ and the explicit 
calculation requires knowing the
correlated (or defect) wave-function for each state-label i.
To obtain the reaction matrix $K$, the inverse scattering method used 
here does not require an explicit
calculation of the correlated  wave-functions. With $K$ known they are however 
obtained from the relation
\begin{equation}
\Psi= \Phi+{Q\over{e}}K.
\label{psi}
\end{equation}

\begin{figure}
\centerline{
\psfig{figure=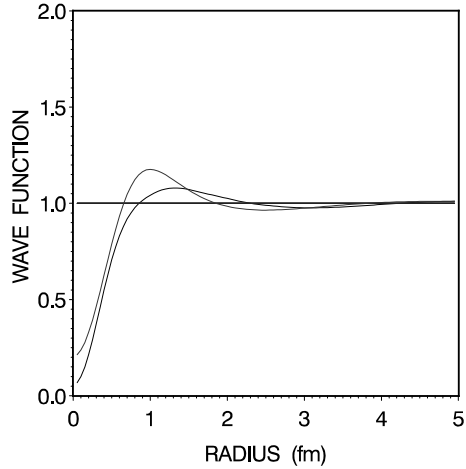,width=7cm,angle=0}
}
\vspace{.0in}
\caption{
The straight line is the uncorrelated wavefunction $\Phi$ at $k=0$.
The lower curve shows the correlated 
$^{1}S_{0}$ while the upper is the correlated $^{3}S_{1}$ wavefunction
$\Psi$
for a relative momentum
$k=0$, a  center of mass momentum $P=0$ and cut-off $\Lambda=9.8
fm^{-1}$.
Note the 'healing'.
For the singlet case this gives a $\kappa=.021$ and for the triplet one
gets $\kappa=.029$
For small radius $\Psi \rightarrow 0$ and
this is evidence of a short-ranged repulsion.
}
\label{wave9.8}
\end{figure}

\begin{figure}
\centerline{
\psfig{figure=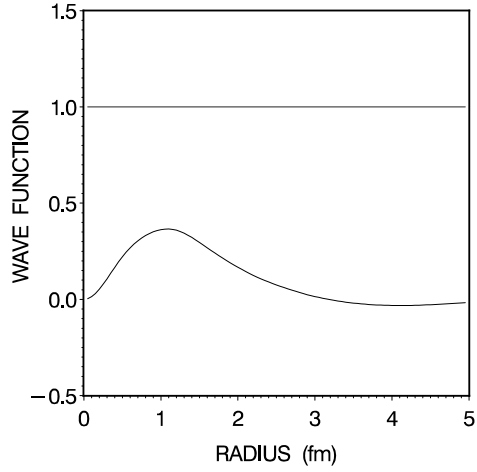,width=7cm,angle=0}
}
\vspace{.0in}
\caption{
The SD defect wavefunction is shown
for a relative momentum
$k=0$ and center of mass momentum $P=0$. The corresponding
$\kappa=.21$. 
}
\label{waveSD}
\end{figure}

\begin{figure}
\centerline{
\psfig{figure=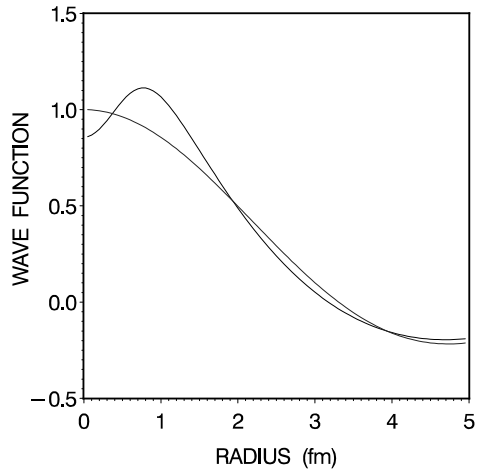,width=7cm,angle=0}
}
\vspace{.0in}
\caption{
The correlated and uncorrelated $^{3}S_{1}$ wavefunctions 
for a relative momentum
$k=0.95 fm^{-1}$, center of mass momentum $P=0$ and cut-off
$\Lambda=9.8$. 
Comparison with the triplet case for $k=0$
in Fig. \ref{wave9.8} shows that the wavefunction at $r=0$ is much
larger, evidently because of the momentum dependence of our two-body
potential. Note the 'healing'. Here $\kappa_{ss}=0.05$
i.e. larger than for $k=0$. 
}
\label{wavetriplet95}
\end{figure}

\begin{figure}
\centerline{
\psfig{figure=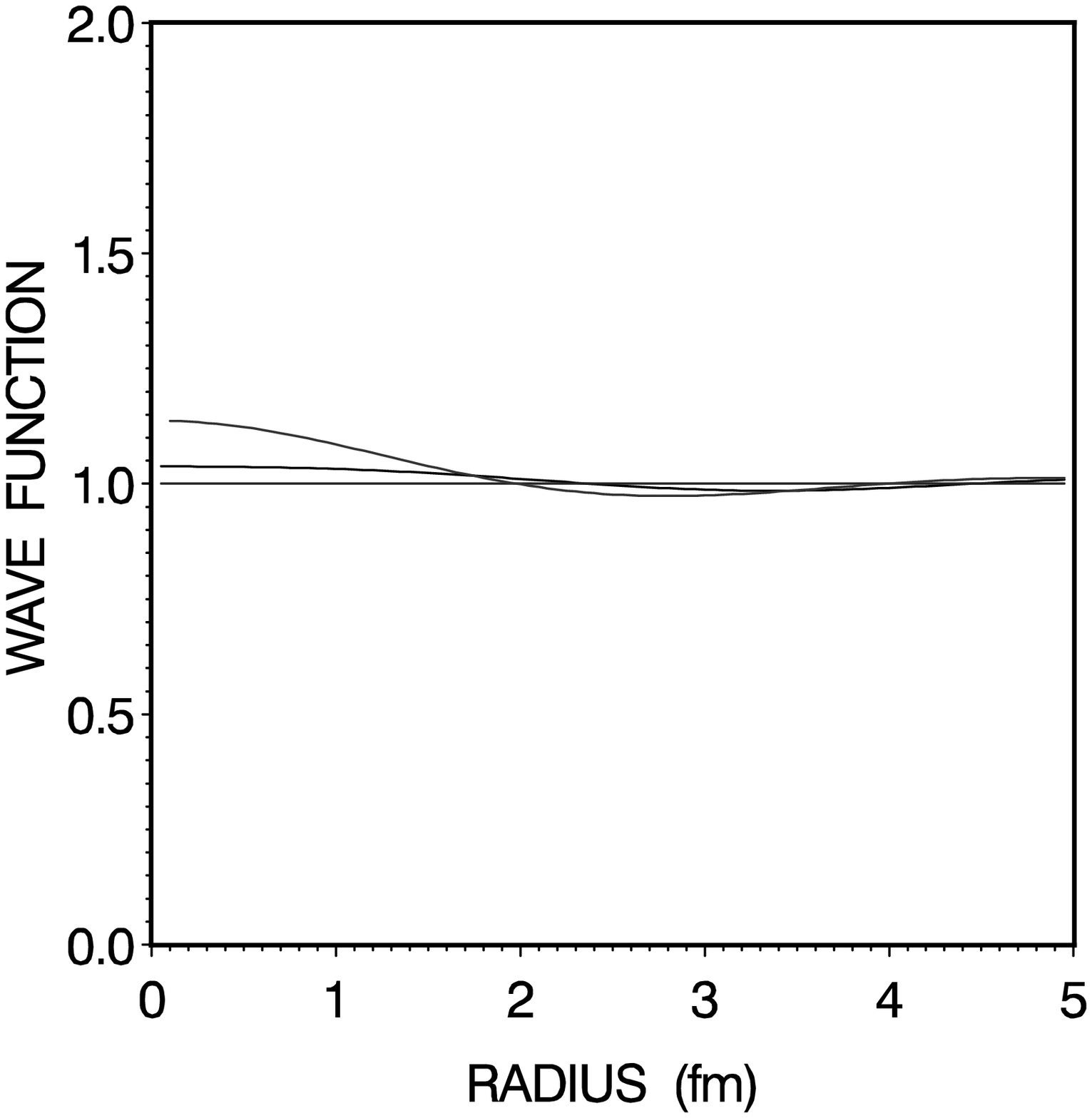,width=7cm,angle=0}
}
\vspace{.0in}
\caption{
The straight line is the uncorrelated wavefunction $\Phi$ at $k=0$.
The lower curve shows the correlated  $^{1}S_{0}$ and the upper the
correlated $^{3}S_{1}$ wavefunction $\Psi$
for a relative momentum
$k=0$, center of mass momentum $P=0$ and $\Lambda=2.0$. 
Compare with the singlet case
in Fig. \ref{wave9.8} for $\Lambda=9.8$.
Here $\kappa=.015$ for the $^{1}S_{0}$ less than the value for
$\Lambda=9.8$ consistent with the independently
calculated average value of $\kappa$ for  shown in the Table
below. 
Compare also with the triplet case
in Fig. \ref{wave9.8} for $\Lambda=9.8$. In this case
$\kappa_{ss}=.013$
There is no evidence of a short-ranged repulsion for this value of
$\Lambda=2.0$.
}
\label{wave2}
\end{figure}

Fig. \ref{wave9.8}  shows  the results for the singlet
and triplet interactions respectively at normal nuclear matter density.
Note the relatively larger defect
in the triplet case for $r \sim 1 fm$. It is due to the tensor-force. 
For the triplet case we also show
the S-D defect wavefunction in Fig. \ref{waveSD}. 
For $k=0$ the unperturbed functions $\Phi=1$
are shown by the straight lines and the well-known
healing is evident. Fig. \ref{wavetriplet95} shows the dependence on
relative momentum when compared with the $^{3}S_{1}$ case in 
Fig. \ref{wave9.8}. Note
especially the difference at radius $r=0$ which is an indication of 
the momentum dependence (non-locality) of our separable interaction 
weakening the short-ranged repulsion  with increasing momentum.
The Figs \ref{wave9.8} to \ref{wavetriplet95}  
are with cutoff $\Lambda=9.8$. 
Our results with $\Lambda=2.0$ are shown in Fig. \ref{wave2}.

One finds in Fig. \ref{wave2} 
that the short ranged repulsive effects are completely absent
in the correlated wave-functions for the cutoff $\Lambda=2.0$. 

The correlated wavefunctions and (more relevant) the defect wavefunctions
and $\kappa$'s depend on the center of mass 
and relative momenta.
In the Table below we show averaged values of $\kappa$ 
calculated from \cite{mos60,hsk93} 
\begin{equation}
\kappa_{i}=
{{1 \over 2}}{\partial P_{i}\over{\partial U}}
\label{kappa-ave}
\end{equation}
where $P_{i}$ is the potential energy for state $i$.
The derivative is performed numerically
by differentiating the potential energy $P_{i}$ with respect to
increments $\pm 0.1 MeV $ in the selfconsistent hole states $U(k)$ for
$k < k_{F}$.

\begin{tabbing}
\hspace{2.6 in} TABLE of $\kappa$'s\\
\\
\hspace{1.5in} $k_{F}=1.35$ \hspace{2.3in} $k_{F}=1.7$ \\
\\
State\hspace{0.2in}\= $\Lambda=9.8$\hspace{0.2in}\=
$\Lambda=4.0$\hspace{0.2in}\= $\Lambda=2.6$ \hspace{0.2in}\=
$\Lambda=2.0$\hspace{0.2in}\= $\Lambda=1.6$ \hspace{0.5in} \=
$\Lambda=9.8$ \hspace{0.2in} \= $\Lambda=2.6$\\

$^{3}S_{1}$ \> $0.109$   \> $0.103$ \> $0.080$       \> $0.062$ \> $0.041$
\> $0.085$ \>$0.045$\\

$^{1}S_{0}$ \> $0.016$   \> $0.016$ \>$0.013$       \> $0.012$ \> $0.011$
\> $0.012$ \>$0.006$\\

$^{3}P_{2}$ \> $0.009$   \> $0.009$ \>$0.007$       \> $0.005$ \> $0.003$ 
\> $0.011$ \>$0.006$\\

$^{1}P_{1}$ \> $0.006$   \> $0.006$ \>$0.005$       \> $0.004$ \> $0.002$ 
\> $0.008$ \>$0.005$\\

$^{3}D_{2}$ \> $0.005$   \> $0.005$ \>$0.004$       \> $0.004$ \> $0.002$ 
\> $0.003$ \>$0.004$\\

$^{3}P_{1}$ \> $0.024$   \> $0.021$ \>$0.014$        \> $0.009$ \> $0.004$ 
\> $0.039$ \>$0.015$ \\

$^{1}D_{2}$ \> $0.002$   \> $0.002$ \> $0.001$        \> $0.001$ \> $0.001$ 
\> $0.003$ \>$0.001$\\

$^{3}P_{0}$ \> $0.002$   \> $0.002$ \>$0.001$       \> $0.001$ \> $0.001$
\> $0.002$ \>$0.001$\\

Total        \> $0.175$   \> $0.163$ \>$0.124$     \> $0.095$    \>$0.061$
\> $0.172$ \>$0.088$
\end{tabbing}

The Table shows that the averaged $\kappa$'s 
decrease with
the cutoff $\Lambda < 4.0 fm^{-1}$ consistent with the  Figs
\ref{wave9.8}
and \ref{wave2} where the correlated wave functions show a very large
$\Lambda$ dependence  especially at \it small radii. \rm 
The argument that $\kappa$ is closely associated with the saturation
property is substantiated by
Fig. \ref{satfig4} that shows the nuclear matter saturation to decrease
for $\Lambda < 4 fm^{-1}$ and this is consistent with Fig. \ref{satfig1}
that shows the dispersion effect to decrease with $\Lambda < 2.6
fm^{-1}$.  

The Table above also shows $\kappa$'s for twice nuclear matter density
($k_{F}=1.7 fm^{-1}$). It is especially noticable that the $\kappa$ for
the $^{3}S_{1}$ state and $\Lambda=9.8$
has decreased by $\sim 20\%$ relative to its value
at normal nuclear matter density. For $\Lambda=2.6$ it has decreased by
$50\%$. This has to be attributed to the
larger effect of the Pauli-blocking cutting off the low-momentum
component of the $SD$ excitations. To substantiate this we show Fig.
\ref{wavetriplet3} to be compared with the $^{3}S_{1}$ curve in 
Fig. \ref{wave9.8}. The
correlations around $r=1 fm$ have decreased by about $10 \%$. It is
however also seen from the Table
that for $\Lambda=9.8fm^{-1}$
the total $\kappa$ at $k_{F}=1.7fm^{-1}$ is practically unchanged
from its value at $k_{F}=1.35$, but for $\Lambda=2.6$ it has decreased
by $\sim 40 \%$. We remind that
$\kappa$ is in fact proportional
to both $\rho$ and to $I_{w}$ so that unless
$I_{w}$ decreases with density as seems to be the case for the
$^{3}S_{1}$ state one does indeed expect $\kappa$ to increase with
density as is also seen for some but not all of the other states.
It may be of interest to note that calculations with the Hamada-Johnston
potential showed a substantial increase of $\kappa$ with density \cite{hsk69}
increasing from
$\sim .24$ at $k_{F}=1.4$ to $\sim .40$ at $k_{F}=1.8$ 
a reflection of the fact that the short-ranged structure of this 
potential is quite different.
(Table 10 in ref \cite{hsk69} shows $I_{w}$ as a function of $k$ and
$k_{F}$). It can also be pointed out that in the Separation
Method\cite{mos60}  the wound-integral is calculated from the "free"
short-ranged correlations and therefore independent of density. 

We like to emphasize that the density \it dependence \rm
of $\kappa$ (and the dispersion correction)  is
of importance in determining not
only the saturation density but also the compressibility.
Likewise we like to point out that 
the dispersion correction  to the energy
(eq. (\ref{dispE})) is proportional not only to
$\kappa$ but also
to the excitation $\Delta U$. 

If $\Lambda$ decreases we have in fact the following scenario.
The short-ranged correlations decrease. 
As a consequence the wound-integral and $\kappa$ decreases. 
Another consequence is that
the excitations to higher
energies where the mean field is less attractive or even repulsive
are suppressed AND the available phase-space is cut off at the
higher end to contribute to this suppression. At the lower end it is the
$Q$-operator that cuts off.
One therefore finds that both  $\kappa$ 
and   $\Delta U$  decrease with $\Lambda$ as a consequence of the
decreased correlations.
So the result is that the dispersion $\Delta E_{disp}$ in eq.
\ref{dispE}
decreases not only because of  a decrease in $\kappa$ 
but also because of  a smaller $\Delta U$. 
In fact, if $\Lambda$ is decreased to $k_{F}$ there are no excitations as they
are completely suppressed by the Pauli-operator and $\Lambda$ and 
the dispersion
correction will be zero. This is clearly seen in Fig. \ref{satfig1} .

\begin{figure}
\centerline{
\psfig{figure=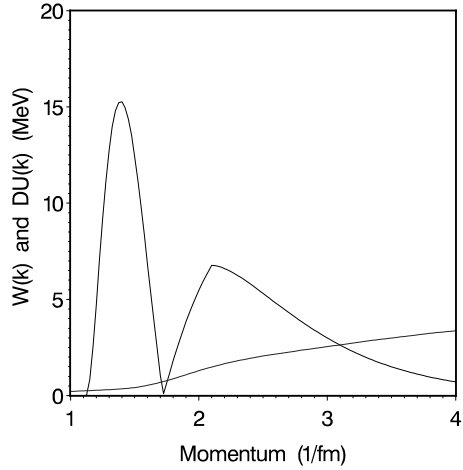,width=7cm,angle=0}
}
\vspace{.0in}
\caption{
The defect wavefunction $W(k)=(\Psi(k)-\Phi(k))^{2}$ is shown together with
$DU(k)=U(k_{1}')+U(k_{2}')-U(k_{1})-U(k_{2})$, 
the latter in units of $\hbar^{2}/2m$. 
The convolution of these functions gives a dispersion-correction of $1.7
MeV$. The contribution to this dispersion from momenta $k<2 fm^{-1}$ is
$0.08 MeV$. Compare with Fig. \ref{wavek2} for $\Lambda=2$.
Here $\Lambda=9.8 fm^{-1}$ and $P=1.5 fm^{-1},k=0.45 fm^{-1}$.
}
\label{wavek9.8}
\end{figure}

\begin{figure}
\centerline{
\psfig{figure=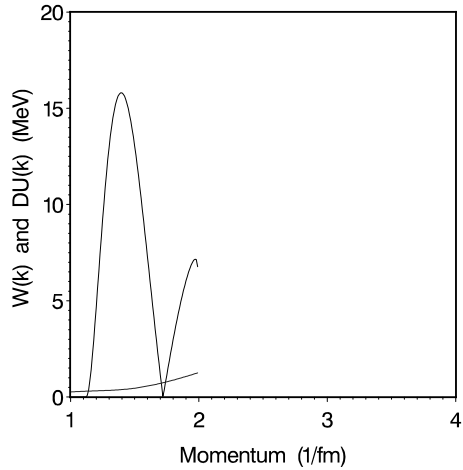,width=7cm,angle=0}
}
\vspace{.0in}
\caption{
Similar to Fig. \ref{wavek9.8} except that $\Lambda=2.0 fm^{-1}$. The
dispersion correction is here obtained to be $0.10 MeV$ nearly the same
as the contribution to dispersion for $k<2$ in Fig. \ref{wavek9.8}.
}
\label{wavek2}
\end{figure}

\begin{figure}
\centerline{
\psfig{figure=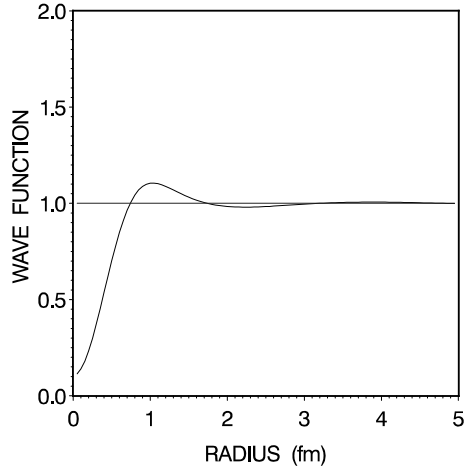,width=7cm,angle=0}
}
\vspace{.0in}
\caption{
The correlated and uncorrelated $^{3}S_{1}$ wavefunctions 
for a relative momentum
$k=0$, center of mass momentum $P=0$ and cut-off $\Lambda=9.8 fm^{-1}$
at twice normal nuclear matter density.
Comparison with the $^{3}S_{1}$ curve in 
Fig. \ref{wave9.8} shows a slight decrease 
in correlations at this higher density. This would explain the decrease in
$\kappa_{^{3}S_{1}}$ with increased density.
}
\label{wavetriplet3}
\end{figure}

Figs \ref{wavek9.8} and \ref{wavek2} further illustrate the situation.
Both Figures are practically identical for momenta $k<2 fm^{-1}$.
From the Table it is seen that the $\kappa$ for the $^{1}S_{0}$ state is
only slightly smaller for $\Lambda=2.0$ than for $\Lambda=9.8$ but the
dispersion corrections are quite different because of the different
cutoffs.

An additional factor to consider in this discussion 
is that the correlations in the medium
i.e. the $\kappa$'s also depend on the chosen mean field that of course
is the origin of the off-shell scatterings.
This is well-known and exemplified by the different $\kappa$'s obtained
with the "continuous" and "standard" choices  of single particle
energies.( See e.g. ref \cite{mac89}).
One concludes that  any discussion of the saturation has to involve
the definition of the mean field. It does not only involve
the correlations.

Higher order correlation effects are discussed in Sect. IV.

It is noted that 
our results for $\kappa$'s are slightly larger than those reported by 
Machleidt \cite{mac89}. He  reports a total $\kappa=0.125$ for the 
Bonn-B potential
with a continuous spectrum. Our larger value 
of $\kappa=0.175$ is consistent
with the larger (more effective)
saturation obtained with our potential\cite{kwo95}.
The methods for calculating the two $\kappa$'s are however not the same.
Machleidt (presumably following ref \cite{haf70}) 
calculates $\kappa$ from the wound-integral by eq (\ref{kappa}) 
at some averaged center of mass and relative momenta, while our averaged
$\kappa$'s are calculated from eq (\ref{kappa-ave}).

The $\kappa$ does depend on the strength of the tensor-force.
We find for comparison a $\kappa=.196$ when using the Bonn-C deuteron
wavefunctions which gives a $P_{D}=5.6$  and $\kappa=.162$ with the
Bonn-A having $P_{D}=4.4$ while Bonn-B has a $P_{D}=5.0$ with
$\kappa=.175.$

\begin{figure}
\centerline{
\psfig{figure=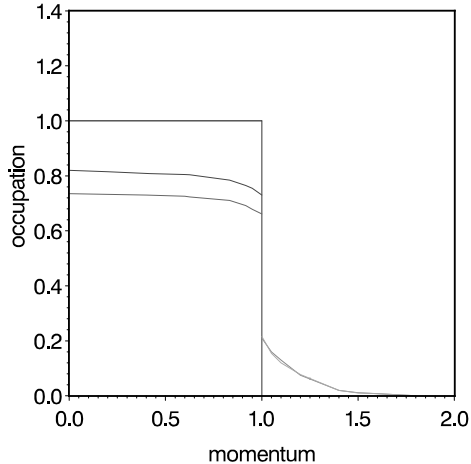,width=7cm,angle=0}
}
\vspace{.0in}
\caption{
Occupation numbers $\rho(p)$ at normal (upper curve) and twice (lower
curve)  normal nuclear matter
density. The momentum is in units of $k/k_{F}$ and $k_{F}=1.35 fm^{-1}$.
The uncorrelated $T=0$ distribution ("square curve") is also shown for
reference.
}
\label{occ}
\end{figure}

The $\kappa$ is a measure of the correlations in nuclei and the
probability of nucleons being scattered out of the fermi-sea.
This depletion can also be caculated from the spectral function
$S(p,\omega)$ with $$\rho(p)={1\over{2\pi}}\int S(p,\omega)d\omega,$$
with $\kappa=1-\rho(p)$.
This has not been done here but
a result of such a calculation is shown in Fig. 
\ref{occ} taken from ref.
\cite{hska92}. It is rather typical of several such calculations found
in the literature that give an occupation at or near 0.8 agreeing with
the measurements quoted below. It is only slightly smaller at the fermi
surface. At twice the
density the Figure shows however a significant decrease in occupation i.e. an 
increase of $\kappa$. This is contrary to the result shown in the Table
above. This increase in $\kappa$ is however not as large as quoted above
for the Hamada-Johnston potential that gave $\kappa=.4$ at this density.
It was already pointed out above that the density-dependence of $\kappa$
is important not only in determining saturation but also compressibility
and it also has consequences for astrophysical theories.
Measurements of spectroscopic factors in $^{208}Pb$ finds depletions of
$0.22 \pm 0.02 \pm 0.06$ for deep-lying states \cite{bat}, that within
error-bars agrees with our result. 

An apparently related quantity (the per-nucleon probability for
two nucleon Short Ranged
Correlations) is measured in recent  
experiments \cite{egi06}  to be 
0.15,0.19, and 0.23  for $^{4}He$, $^{12}C$, and $^{56}Fe$ respectively.
These values are larger than the $\kappa$'s shown in our calculations
above. This would require even stronger correlations.
The exact interpretation of the experiments may still have to be
clarified however. 
The $\kappa$ for the $^{3}S_{1}$ state is seen above to be $\sim 6$
times as large as the $\kappa$ for the $^{1}S_{0}$ state. This is 
consistent
with recent experiments \cite{pia06}.

\section{Tensor correlations}
The numerical results above  agree with the well-known fact that a dominant
contribution to the dispersion-term and saturation comes  from
the tensor-component.(see e.g. ref \cite{mac89})
The short ranged correlations in the
$^{1}S_{0}$ state and in the $^{3}S_{1}$ state with the tensor force
swithched off contribute much less to the dispersion correction 
as shown by the numerical results
in Figs \ref{satfig1} and \ref{satfig2}.
This situation leads however to a big dilemma because of the experimental and
theoretical uncertainty regarding the strength of the 
tensor-force at short distances, which is an important  factor in
calculating the $D$ state probability $P_{D}$ in the deuteron. 
The long-ranged part is of course 
accurately determined by the pion-exchange contribution.
While the deuteron quadrupole-moment $Q$ and the
asymptotic ratio $\eta =D/S$ are well known experimentally the $D$-state
probability $P_{D}$ is only known approximately, being somewhere between
4 and 7 $\% $.  This dilemma was already emphasized by Machleidt 
who defined three phase-shift equivalent potentials Bonn-A, Bonn-B and
Bonn-C having different tensor-strengths with values of $P_{D}$
being $4.4,5.0$ and $5.6$ respectively.\cite{mac89}
These potentials
gave very different saturation properties. This was also shown to be the
case with the inverse scattering potentials used in the present
work.\cite{kwo95} 
For a fixed range of the
deuteron wavefunctions the saturation energy and density will decrease
with the D-state probability $P_{D}$.\cite{mac89,kwo95}
The relation between saturation and the uncertainty of
the deuteron wavefunction was further illustrated by the class of
potentials named FBS in ref \cite{kwo95}. These were derived from
deuteron wavefunctions constrained by the deuteron-data ($\eta$ and Q)
but with  longer tails in momentum space. Parametrical fits to 
the $d(e,e'p)n$ experimental
data by Bernheim \it et al\rm \cite{ber81} available for $k< 1.7
fm^{-1}$ were made.
Increasing the range of the
wavefunctions i.e assuming a longer tail in momentum space (in the region
not determined experimentally),
but keeping  $P_{D},Q$ and $\eta$ fixed, will 
have an effect similar to that of decreasing $P_{D}$.
\cite{kwo95}
The $P_{D}$ is however mainly
determined by the wave-function in the region around  $k=2 fm^{-1}$ not
quite
available from the Bernheim \it et al \rm data that were limited
to $p<335 Mev/c$. The need to go to higher energies to fully explore the
$D$-state distribution was also stressed by Bernheim \it et al \rm.

More recent efforts to explicitly determine the tensor-component in the
NN-interaction were made by measuring elastic electron-deuteron
scatterings for momentum transfers up to $Q^{2}=1.7 (GeV/c^{2})$.
\cite{abb00} Of particular interest here are the $T_{20}$ data which
should be closely related to the tensor-force. 
Comparisons of these scattering data with relativistic
calculations\cite{car99} as well as QCD\cite{bro92,kob94} show the
former to agree fairly well with the data while not the latter.
It is however well recognised that the meson theoretical calculations
are hampered by incomplete knowledge of meson exchange currents (MEC).

The question of the strength of the tensor-force and therefore its role
in the problem of nuclear saturation is therefore still unresolved at
this time. It may in fact be the most important unsolved problem in
nuclear (many body) physics.

\section{Higher Order Corrections}
The results presented here were all obtained in the Brueckner
approximation given by eqs. (\ref{K},\ref{U},\ref{E}).
Higher order  rearrangement corrections were already discussed and
estimated by Brueckner and Goldman \cite{bru60}. 
  and in many subsequent
works e.g. refs. ( 
\cite{hsk93,hsk66,hsk73,hsk92,bal89}).
More recent but related work has concentrated on corrections due to
spectral selfconsistency and inclusion of hole-hole ladders. 

\begin{figure}
\centerline{
\psfig{figure=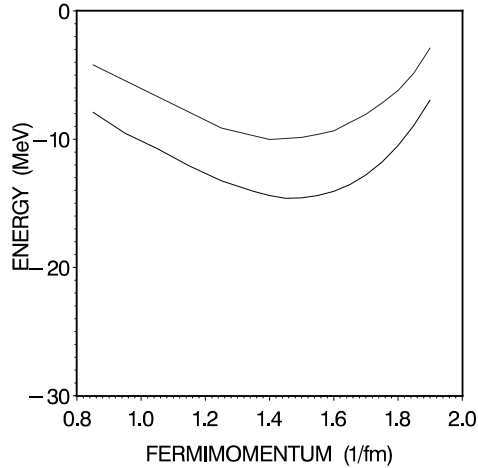,width=7cm,angle=0}
}
\vspace{.0in}
\caption{
The lower curve is the result of a Brueckner calculation (i.e. with
particle-particle ladders only) of the total
energy as a function of density (Fermimomentum). The lower curve includes
hole-hole ladders perturbatively by changing the Pauli-operator in eq.
\ref{K}.
}
\label{holehole}
\end{figure}

The Brueckner $K$-matrix includes only particle ladders while hole-hole
ladders are also naturally included in Green's function techniques.
It was shown that, in the quasi-classical limit, the Green's function mean
field $Re\Sigma^{+}$= $U+ReU^({2})$ with $U^({2})$ the Brueckner second order
rearrangement\cite{hsk93}. In this limit hole-hole ladders can therefore
be included simply by including $U^({2})$ in the calculation of the total
energy $E$ from eq. (\ref{E}). 
Our result of such a calculation is shown
in Fig. \ref{holehole}. 
This is a perturbative calculation.
Higher order is well known to be divergent.
The lower curve is a
Brueckner calculation while 
the upper includes the hole-hole ladders.
It is seen to decrease the binding by $\sim 4 MeV$. 

The spectral broadening was first calculated from the
second order Brueckner rearrangement energy \cite{hsk66} .
It was then also concluded
that the width stems from the long-ranged part of the interaction. In
a subsequent work  this broadening of the spectral function was included
selfconsistently in a nuclear matter calculation \cite{hsk05}. This was done 
using Green's function techniques. Only the long-ranged part of an
interaction was used so that any effect of shortranged correlations was
not included. It was found that the spectral broadening
increased the binding energy and also slightly the saturation density.
While the long-ranged correlations give a spectral broadening but
essentially no depletion 
the shortranged correlations do cause depletion.

In ref \cite{hsk73} the
effect of this depletion on the selfenergy was calculated from the
Brueckner
third order rearrangement energy. This is a short-ranged effect and it 
results in a "renormalisation" of the
mean field to get \cite{hsk75} $$U(k) \rightarrow(1-\kappa U(k)).$$ 
When reinserted in eq. (\ref{K}) the result is an increased binding as a
result of decreased dispersion-correction by eq. (\ref{disp}). The
saturation density is also increased in such a calculation. 
The Brueckner total energy given by eq. (\ref{E}) is however a
quasi-particle approximation summing over occupation numbers 0 and 1. 
It was shown in ref. \cite{hsk92}
that this can overestimate the binding relative to  a summation over
spectroscopic strengths as in Koltun's sum rule.

The above results  indicate and it has been shown repeatedly that, if one goes 
beyond the quasi-particle approximation,
spectral selfconsistency should be invoked . 
The \it ab initio \rm quasi-particle (Brueckner)
calculations have been extended by several authors
utilizing Green's function formalisms
thereby including corrections due to the broadening and depletion
of the spectral functions as well as the hole-hole ladders.
With present-day computer-power this
is now feasible \cite{boz06,muth03,boz99,boz02}.  Consequently the width- and
depletion-effects calculated above are now included selfconsistently as
they are imbedded in the spectral functions.
Comparison with Brueckner calculations shows a decrease in binding very
similar to what is seen in Fig. \ref{holehole}\cite{muth03}.
This can be regarded as a 
relatively small correction and 
shows near cancellation of the different effects mentioned above.

This extension of Brueckner's original theory does however not address
other  higher order effects that may affect two-body correlations and
saturation. One issue is for example alternative insertions in particle
lines that would affect the $\Delta U$ in eq. (\ref{disp}) for the
dispersion correction. A rigorous treatment of these insertions was made
by Bethe who realised that they should be treated as 3-body collisions
\cite{bet65}. It has however been claimed by Song \it et al \rm
\cite{son98} that all three-hole line contributions are essentially
included if the continuous choice of the spectrum is used as we have
also done here.

We believe that there are still uncertainties relating to both the
NN-interaction at short distances including the tensor component and in
the neglect of higher order corrections in the mamy-body theory. Another
issue is of course 3-body forces that we have not included in this work.

The present evaluation of corrections discussed above does however 
not alter our main conclusions regarding saturation
and the cut-off parameter $\Lambda$ although
it can not be ruled out that our values of $\kappa$ can
be slightly changed by improved computing
techniques \cite{fri02}. This should be investigated. It was however
shown that the Brueckner result in eq. \ref{kappa-ave} agrees with the
Green's function result if using the EQP- (Extended Quasi-Particle-)
approximation for the spectral function. (See eq. (66) in ref.
\cite{hsk93}).

\section{Summary and Conclusions}
 Many-body calculations with realistic nuclear forces have a long
 history.  Various methods have been used:
 Brueckner, $e^{S}$, coupled cluster, HNC and
 maybe others. These more or less agree in that realistic 2-body forces
 and a plausible 3-body force can provide reasonable saturation of nuclear
 matter and experimental fits to the lightest nuclei. The effective
 "in-medium" 2-body force is density-dependent and the 3-body force can
 also be regarded as an effective density-dependent 2-body force.
Without any reference to a specific theory one can therefore conclude
that the effective force in the
nuclear medium has to be  density-dependent in order to 
achieve the observed saturation.
The origin of this density-dependence is however only partially understood
theoretically. 

The philosophy behind the $V_{low-k}$ and EFT methods is that the
relative momenta of nucleons in nuclei are  low  and that
the high-momentum components
of the NN-interaction 
therefore should be allowed to be integrated out since they are not well
known anyway. The result is 
a renormalised "smooth" effective force that may even be treated in low
order perturbation theory. 

The $V_{low-k}$ effective force does in itself not provide saturation
and has to be supplemented with a $3N$ force of not well defined
origin.\cite{bog05} 
The effective force generated by EFT-methods does explicitly generate a
3-body force but apparently not sufficiently strong to provide
saturation. 

In Brueckner theory the effective force is obtained by replacing the 
free space propagator $1/e_{0}$ in the  
nuclear medium by $Q/e$. 
The $Q$ operator as well as the mean field included in 
the energy-denominator $e$
contribute to the density-dependence.
The momentum-dependent (non-local) mean field results in 
off-shell scatterings in the many-body medium which together with 
short-ranged correlations  yields a
dispersion term (eq. (\ref{disp})) which is  density-dependent. This is a
major contributor to the density-dependence of the 
effective force in Brueckner theory.
The range of the
correlations contributing to this density-dependence 
is typically long (in momentum space) compared to 
the fermi-momentum and to the typical cut-offs used to produce the renormalised
low-momentum effective forces. The question addressed in this paper is
which effect  the cut-off has on the density-dependence and consequently 
the saturation and the compressibility of nuclear matter. 

The study is  
done with a separable potential derived by previously
published inverse scattering methods. \cite{kwo95}
The input  consists of the scattering phase-shifts and the
deuteron data. The effect of the cut-off of the high momentum
phase-shifts on the effective interaction was shown in a previous 
publication \cite{hsk04} and  the similarity with $V_{low-k}$ was shown.
In the
present work 
a primary interest has been the dispersion effects.

The effect of the cut-offs on the dispersion correction
was shown in Figs 1-2. It is found that at
normal saturation density a cut-off larger than about $3fm^{-1}$ is
necessary to include the full effect of the correlations on dispersion and 
the saturation property of the force.

We reaffirm the importance of 
tensor-correlations in Section III.
They are not
fully defined by the scattering phase-shifts but the deuteron
properties are  important inputs. It was already shown by 
Machleidt \cite{mac89} that deuteron
wave-functions constrained by the known deuteron quadrupole moment and the
asymptotic D/S ratio but having different unknown D-state probabilities give  
different saturation results.\cite{mac89} This was confirmed by the inverse
scattering results 
of Kwong and K\"ohler \cite{kwo95},
who further investigated the effect on saturation 
using three additional model wave-functions  for the deuteron. 
The D-state admixture is largely unknown for momenta above $\sim 2
fm^{-1}$. This leaves a major uncertainty in the theory of saturation.

It is now generally accepted that an important 
part of the density-dependence comes from 3-body
forces.
It seems  however that at least part of the unknown factor regarding
saturation rests with two-body 
correlations especially in the $^{3}S_{1}$ channel.
Because of this uncertainty it
seems reasonable to look for alternatives. One such is the
Moscow-potential of Neudatchin \it et al \rm \cite{neu75,mos05} that has a
strong short-ranged attraction that changes the phase-shifts by an
unobservable amount of $\pi$ radians resulting in a node in the relative
wavefunction. This leads to a strongly correlated
short-ranged system resulting in larger wound-integrals and dispersion
corrections. If the tensor force is not sufficient to give sufficient
saturation this force may be the right answer to the saturation problem.

We have here only been concerned with the effect of correlations and
momentum cut-offs in nuclear matter calculations.  A somewhat different
problem is that of the effect of cut-offs for nuclear structure
calculations.
It is however shown
in Fig. \ref{satfig4} that for densities below the saturation density,
the total binding energy is practically independent of cut-off at least
for $\Lambda > 2.6 fm^{-1}$.
This suggests that short-ranged correlation effects may be of less importance
in finite nuclei as long as saturation is not an issue, i.e. if the density
distribution is constrained by fixing the nucler size.
If not, the calculation would result in a finite
nucleus collapsing to  a too small radius.

\begin{figure}
\centerline{
\psfig{figure=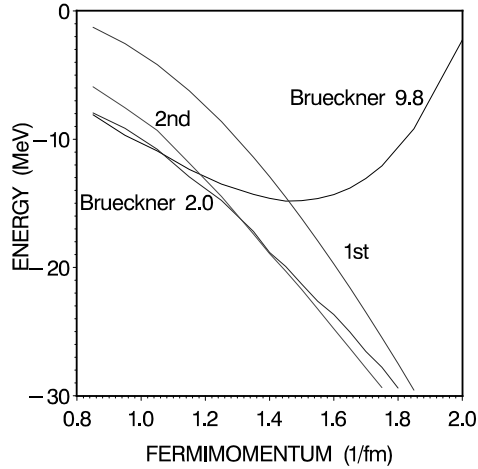,width=7cm,angle=0}
}
\vspace{.0in}
\caption{
 Brueckner calculations of the binding energy per particle for
$\Lambda=9.8$ and $2.0 fm^{-1} $ together with first and second order 
calculations with an interaction defined with a cut-off $\Lambda=2.0$.
}
\label{br1st2nd}
\end{figure}

One motivation to develop low momentum effective interactions has been
that it might be used in low order perturbation theory as
opposed to the more traditional many-body techniques developped for
strongly interacting media. \cite{bog05}
In Fig. \ref{br1st2nd} are shown four different curves. 
These are results of two Brueckner, one first and one second order
calculation of nuclear matter binding energies, the latter with a cut-off
$\Lambda=2.0$. 
As expected there is no sign of saturation for $\Lambda=2.0$. This 
was already seen to be the case for
the Brueckner calculation with $\Lambda=2.6 fm^{-1}$ in Fig.
\ref{satfig4}. However, the
second order result shows a considerable improvement over the first
order
at densities below saturation approaching the Brueckner result with the same
$\Lambda$. 

\newpage


 

 


\begin{thebibliography}{10}
\bibitem{bet36} H. A. Bethe and R.F. Bacher,
                Rev. Mod. Phys.  {\bf 8} (1936) 82.
\bibitem{bro67} G.E. Brown, 'Unified Theory of Nuclear 
                Models and Forces',
		North-Holland Publishing Co. 1967.
\bibitem{jas51} R. Jastrow,
                Phys. Rev. {\bf 98} (1951) 165.
\bibitem{mos60} S.A. Moszkowski and B.L. Scott
	        Ann. of Phys. {\bf 11} (1960) 65.
\bibitem{hsk61} H.S. K\"ohler ,
	        Ann. of Phys. {\bf 16} (1961) 375.
\bibitem{kwo95} N.H. Kwong and H.S. K\"ohler, 
                 Phys. \ Rev. C {\bf 55} (1997) 1650.
\bibitem{hsk04}  H.S. K\"ohler ,
                nucl-th/0511030.
\bibitem{fri02} T. Frick, Kh.Gad, H. M\"uther and P. Czerski
                Phys. \ Rev. C {\bf 65} (2002) 034321.
\bibitem{boz06} P. Boz\'ek, D. J. Dean and H. M\"uther,
                nucl-th/0604003.
\bibitem{dah69} G. Dahl, E. Ostgaard and B. Brandow,
                Nucl. Phys. {\bf  A124} (1969) 481.
\bibitem{hsk69} H.S. K\"ohler ,
                Nucl. Phys. {\bf  A128} (1969) 273.
\bibitem{hsk93} H.S. K\"ohler and Rudi Malfliet,
                Phys. Rev. C {\bf 48} (1993) 1034.
\bibitem{haf70} Michael J. Haftel and Frank Tabakin
                 Nucl. Phys {\bf A158} (1970) 1.
\bibitem{bet71} H.A. Bethe,
                Ann.Rev Nucl. Sci. {\bf 21} (1971) 93. 
\bibitem{mac89} R. Machleidt, Adv. Nucl. Phys. {\bf 19} (1989) 189.
\bibitem{ber81} M. Bernheim, A. Bussi\'ere, J. Mougey, D. Royer, D.
                Tarnowski, S. Turck-Chieze, S. Frullani, G.P. Capitani,
		E. de Sanctis, and E. Jans,
                 Nucl. Phys {\bf A365} (1981) 349.
\bibitem{abb00} D. Abbott \it et al \rm 
                Phys. Rev. Lett. {\bf 84} (2000) 5053.
\bibitem{car99} J. Carbonell and V.A. Karmanov
                The European Physical Journal A {\bf 6} (1999) 9.     
\bibitem{bro92} S. J. Brodsky and J. R. Hiller 
                Phys. Rev.  D{\bf 46} (1992) 2141.
\bibitem{kob94} A. Kobushkin and A. Syamtomov,
                Phys. Rev. D {\bf 49} (1994) 1637.
\bibitem{bru60} K.A. Brueckner,
                Phys. Rev.  {\bf 117} (1960) 207.
\bibitem{hsk66} H.S. K\"ohler ,
                Nucl. Phys. {\bf  88} (1966) 529.
\bibitem{hsk73} H.S. K\"ohler ,
                Nucl. Phys. {\bf  A204} (1973) 65.
\bibitem{hsk75} H.S. K\"ohler ,
                Phys. Reports {\bf  18} (1975) 217.
\bibitem{bal89} M. Baldo, I. Bombaci, G. Giansiracusa, U. Lombardo,
                C. Mahaux and R. Sartor,
                Phys. Rev.  {\bf 117} (1960) 207.
\bibitem{hsk92} H.S. K\"ohler,
                Phys. Rev. C {\bf 46} (1992) 1687.
\bibitem{boz99} P. Bo\'zek,
                Phys. Rev. C {\bf 59} (1999) 2619, Phys. Rev. C {\bf 65}
		(2002) 054306.
\bibitem{boz02} P. Bo\'zek and P. Czerski nucl-th/0212035.,
\bibitem{muth03} T. Frick and H. M\"uther,
                Phys. Rev. C {\bf 68} (2003) 034310.
\bibitem{hsk05} H.S. K\"ohler,
                nucl-th/0509060.
\bibitem{bog05} S.K. Bogner, A. Schwenk, R.J. Furnstahl and A. Nogga
                nucl-th/0504043.
\bibitem{bet65} H. A. Bethe,
                Phys. Rev.  {\bf B138} (1965) 804.
\bibitem{son98} H.Q. Song, M. Baldo, G. Giansiracusa and U. Lombardo,
                Phys. Rev. Lett. {\bf 81} (1998) 1584.
\bibitem{bat}   M.F. Van Batenburg thesis
                http://igitur.archive.library.uu.nl/dissertation/
		1952912/inhoud.htm
\bibitem{hska92} H.S. K\"ohler ,
                Nucl. Phys. {\bf  A537} (1992) 64.
\bibitem{egi06} K.S. Egiyan et al
                Phys. Rev. Lett. {\bf 96} (2006) 082501.
\bibitem{pia06} E. Piasetzky, M. Sargsian, L.Frankfurt, M. Strikman, and
                J.W. Watson '
                nucl-th/0604012
\bibitem{neu75} V.G. Neudatchin, I.T. Obukhovsky, V.J. Kukulin and N.F.
                Golovanova,
                Phys. Rev. C {\bf 11} (1975) 128.
\bibitem{mos05} S.A. Moszkowski, in
Proceedings online of the Conference
on Microscopic Approaches\\ to Many-Body Theory (MAMBT)
in honor of Ray Bishop, Manchester, UK, \\
http://www.qmbt.org/MAMBT/pdf/Moszkowski.pdf, 2005.
\end{thebibliography}
\end{document}